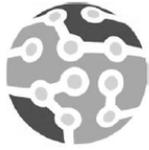

# Interfaz Gráfica para la Gestión SDWN de un Entorno WLAN

Pedro Fortón, José Mª Saldaña, Julián Fernández-Navajas, José Ruiz-Mas

Departamento de Ingeniería Electrónica y Comunicaciones – Instituto de Investigación en Ingeniería de Aragón (I3A)

Universidad de Zaragoza

C/ María de Luna 1, Edif. Ada Byron, 50018 Zaragoza, España.

jsaldana@unizar.es, navajas@unizar.es, jruiz@unizar.es

*Resumen*- En este trabajo se presenta una interfaz gráfica para la gestión de una red de área local inalámbrica que integra un conjunto de puntos de acceso Wi-Fi coordinados. La interfaz interactúa con una aplicación de red que se encarga de realizar un balanceo de carga y una gestión de la movilidad. La aplicación gráfica es capaz de obtener y mostrar la información almacenada en un sistema que incluye un número de puntos de acceso, y la muestra de manera amigable para el gestor de la red. Finalmente, la aplicación permite la gestión remota del sistema, ajustando sus parámetros o realizando traspasos entre puntos de acceso.

*Palabras Clave*- SDWN, Wireless LAN, 802.11, interfaz gráfica

## I. Introducción

En los últimos años se ha experimentado un gran crecimiento en el uso de dispositivos móviles (teléfonos, tabletas, portátiles, etc.), lo que ha llevado a un aumento de la necesidad de permanecer conectados a Internet en todo momento y lugar (casa, trabajo, ocio, etc.). Aunque actualmente el despliegue de redes de datos 3G o 4G está generalizado en muchos países, las redes inalámbricas basadas en IEEE 802.11 [1], conocidas como redes Wi-Fi, siguen siendo usadas con mucha frecuencia por los usuarios, debido a su compatibilidad con diversos dispositivos, su ancho de banda, su presencia en la mayoría de los hogares y, por supuesto, su gratuidad en muchos lugares públicos (bares, restaurantes, hoteles, aeropuertos, etc.).

Actualmente el número de redes Wi-Fi que podemos encontrar en una misma localización es muy alto, compartiendo todas ellas unos recursos espectrales limitados. Esa situación se conoce como Wi-Fi *Jungle* [2], ya que los puntos de acceso (*Access Point*, APs) comparten canales y la interferencia es elevada.

En los lugares en los que es posible coordinar los puntos de acceso, como centros comerciales, empresas o edificios de servicios públicos, se suelen desplegar soluciones denominadas WLAN *Enterprise*, que son redes inalámbricas de clase empresarial que mejoran en ciertos aspectos la versión para uso personal, que se suele denominar SOHO (*Small Office / Home Office*).

Debido a que las soluciones WLAN Enterprise actuales son tecnología propietaria y de alto coste, algunos trabajos han estudiado y desarrollado herramientas de código abierto que realizan ese servicio utilizando hardware de bajo coste, centrándose en la movilidad y el uso eficiente de los recursos inalámbricos. Algunas soluciones se basan en la adaptación de las ideas de SDN (*Software Defined Networks*) a entornos inalámbricos y se suelen denominar SDWN (*Software Defined Wireless Networks*) [3], creado para cada terminal cuando éste accede a la red y al que permanece conectado mientras sea posible.

Gracias a esta abstracción, un mismo punto de acceso físico puede albergar varios LVAP, dando acceso a cada terminal mediante diferente BSSID (*Basic Service Set Identifier*). Todo ello facilita la gestión del controlador a la hora de decidir qué AP da servicio a un determinado terminal o estación (STA), para lo cual debe ser capaz de "trasladar" el LVAP de un AP a otro, siguiendo el desplazamiento del terminal asociado, de una manera transparente para él [4].

La Fig. 1 ilustra el funcionamiento del LVAP: un controlador puede moverlo entre diferentes AP físicos, de forma que la estación siempre recibe las tramas de la misma dirección TA *(Transmitting Address)*, y "piensa" que siempre está conectada al mismo AP. Esto requiere un "parche" en el driver de la tarjeta inalámbrica del AP, para que se modifique TA en función de la STA destino. Es decir, un mismo AP utilizará un BSSID distinto en función de la STA destino.

Dentro del proyecto H2020 Wi-5 se desarrollaron algunas soluciones, llamadas *aplicaciones*, en las que se permitía el balanceo de carga y la gestión de la movilidad de los usuarios en escenarios con un número elevado de AP [5]. Sin embargo, el proyecto Wi-5 no planteó el desarrollo de una interfaz amigable que permitiera la monitorización del estado de la red,

o el análisis de la información generada por el sistema. La información se mostraba mediante trazas en la consola del equipo controlador de la aplicación, lo que limitaba las posibilidades de explotación de los datos generados por el sistema.

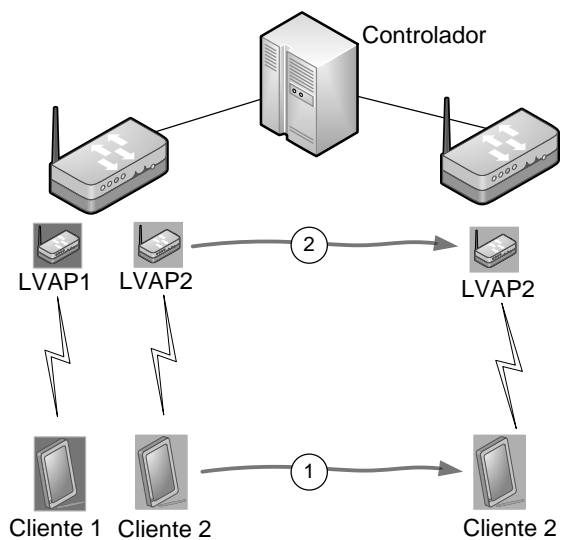

Fig. 1. Traslado de LVAP por orden del controlador.

Partiendo de la necesidad de un mejor uso de los datos generados por el conjunto de aplicaciones disponibles, el presente artículo expone el desarrollo de una interfaz gráfica usable y amigable, que presente los datos en tiempo real y permita la interacción con el sistema. Entre las múltiples aplicaciones de red desarrolladas por el proyecto Wi-5, se ha elegido desarrollar dicha interfaz gráfica para la más completa de todas, *Smart AP Selection*. Esta aplicación combina los resultados de varias aplicaciones desarrolladas durante el proceso del proyecto, y permite una gestión de la movilidad de las STA y el balanceo de carga [6].

## II. DESCRIPCIÓN DEL ENTORNO ODIN WI-5

En esta sección se describirán en primer lugar los elementos que componen el sistema, para posteriormente explicar la aplicación de gestión de la movilidad y balanceo sobre la que se ha añadido la interfaz gráfica.

### A. Elementos del sistema

Se parte de la solución Wi-5 [5], que se desarrolló como una extensión de la plataforma Odin [4], presentada en [7], que se compone de los siguientes elementos (ver Fig. 2):

- El controlador *(Controller)*, implementado como un módulo para el Floodlight OpenFlow Controller (http://www.projectfloodlight.org/), sobre el cual, el grupo de investigación de la Universidad de Zaragoza desarrolló un conjunto de aplicaciones (*Smart Functionalities* o Wi-5 *apps*) que permiten balancear la carga, gestionar la movilidad, obtener estadísticas de red en tiempo real, etc.
- Los AP, que son gestionados por el controlador, y deben realizar traspasos suaves de terminales (*seamless handoff*) mediante el uso de LVAP. Además realizan funciones de monitorización de terminales, tráfico y potencia de señal, y envían la información al controlador, que es quien decide qué hacer en cada situación.

### B. Aplicación de gestión de la movilidad y balanceo de carga

La aplicación *Smart AP Selection* forma parte del Wi-5 *controller*, que está integrado dentro de Floodlight Controller. Esta aplicación utiliza un método proactivo para realizar una gestión de la movilidad de los usuarios (asignar cada dispositivo al AP que resulte óptimo), y simultáneamente un balanceo de carga (conseguir una distribución equilibrada entre los AP) [6].

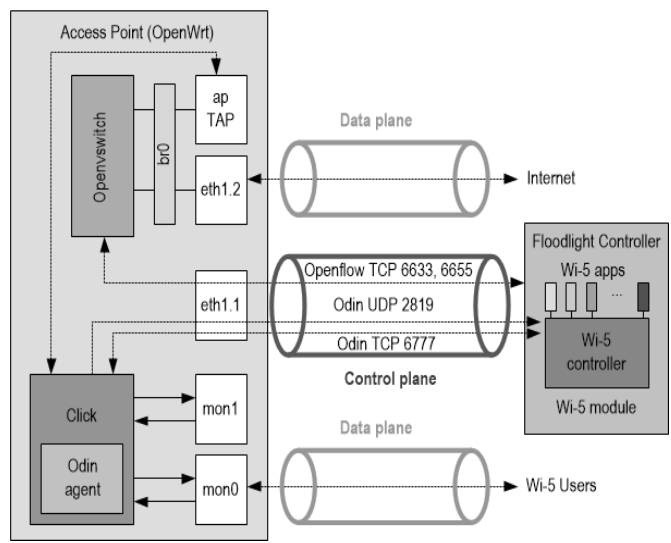

Fig. 2. Elementos del sistema Wi-5.

Mediante el protocolo Odin [4], solicita de manera periódica a los diferentes puntos de acceso de la red un escaneo en sus respectivos canales. La información obtenida en tiempo real permite la asignación al controlador, mediante diferentes algoritmos, la asignación a los AP de las estaciones conectadas a la red, según la posición en la que se encuentren *(mobility management)*, así como distribuir óptimamente el tráfico entre los AP disponibles *(load balancing)*. El tiempo empleado en la iteración principal debe ser lo suficientemente breve como para permitir al controlador tomar decisiones de traspaso entre puntos de acceso para usuarios que caminan. En las pruebas realizadas durante el proyecto Wi-5, se mostró que, para seguir adecuadamente a un usuario máximo, estas decisiones se deben tomar cada 2 segundos como máximo [6].

La función principal de la aplicación realiza, en el siguiente orden, estas acciones (Fig. 3):
1. Inicializa las variables que se utilizarán a lo largo de toda la ejecución.
2. Solicita un escaneo activo a todos los AP conectados al controlador, solicitando su nivel de RSSI *(Received Signal Strength Indicator*, el indicador de fuerza de la señal recibida) recibido de cada una de las estaciones conectadas.
3. Con la información obtenida a través de cada una de las estaciones, la función crea y almacena una matriz de atenuación. Haciendo uso de esta matriz, un algoritmo organizará la red en busca de una configuración óptima.
4. Finalmente, las decisiones del algoritmo se implementarán en la red, realizando los traspasos oportunos. Estos traspasos se consiguen mediante la asignación de un LVAP a un nuevo AP. El LVAP está asociado a cada cliente, y "viaja" con él, de forma que cada cliente se comunica siempre con la MAC del LVAP.

Para calcular la matriz de atenuación (paso 3) después de cada escaneo periódico, se realiza un cálculo en el cual se obtiene la atenuación suavizada (RSSI) para cada par AP-estación [6]:

$$RSSI\ suavizada = \alpha * nuevo\ RSSI +$$
$$+ (1-\alpha) * RSSI\ histórico \quad (1)$$

Una vez finalizado el proceso que calcula la matriz de atenuación, da comienzo el proceso que asigna las estaciones a sus puntos de acceso óptimos (paso 4), teniendo en cuenta el RSSI suavizado y el número de estaciones que hay en cada AP.

## III. DESARROLLO DE UNA INTERFAZ GRÁFICA

En esta sección se explica en detalle el desarrollo y los elementos de la interfaz gráfica que se ha desarrollado, y que permite al usuario interactuar con la aplicación *Smart AP Selection*, para obtener información en tiempo real sobre el estado de la red y permitir una gestión manual de los resursos (canales, traspasos, etc.) que facilite el desarrollo de algoritmos para su automatización.

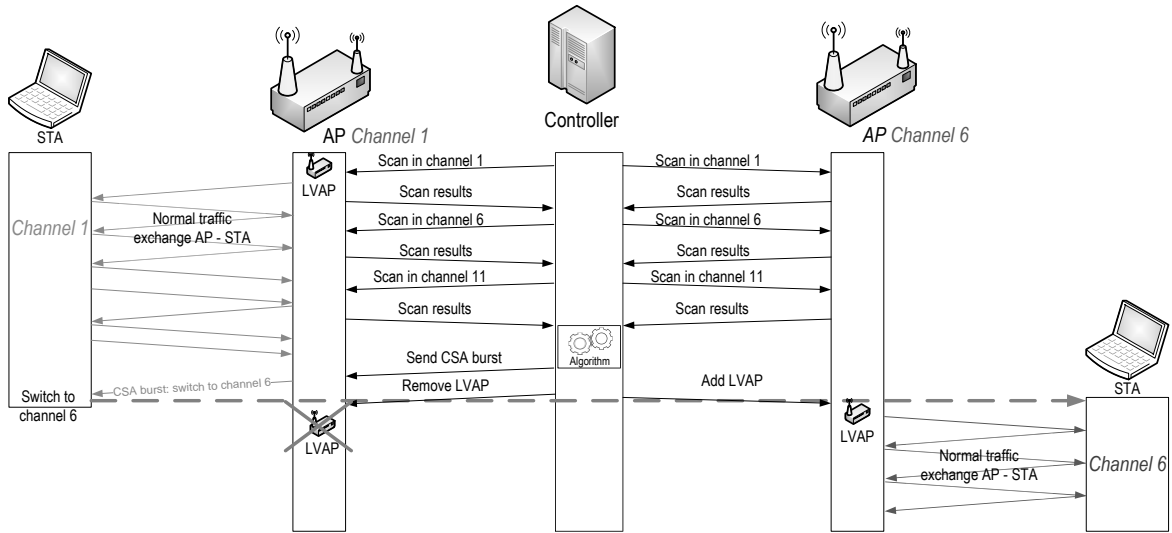

Fig. 3. Esquema de funcionamiento de *SmartAPSelection*.

El código de la interfaz desarrollada se encuentra disponible en el repositorio github del proyecto Wi-5:

https://github.com/Wi5/odin-wi5-gui-client

### A. Arquitectura del sistema

La arquitectura del sistema (Fig. 4) está compuesta por diferentes capas:

- *Odin*. Contiene la aplicación *Smart AP Selection* con las funciones correspondientes de asignación de canales y terminales y traspasos mediante los diferentes algoritmos implementados y los datos recopilados mediante el intercambio de flujos de información entre el controlador y los diferentes AP.
- *Servidor*. Contiene los servicios necesarios para la extracción de datos de la capa Odin y su almacenamiento NoSql y el uso de sus funciones. Además dispone de un servidor web para interactuar con ella.
- *Cliente* (interfaz gráfica propiamente dicha), independiente del servidor, que ataca la capa servidor mediante peticiones HTTP a su servidor web.

Por otra parte, la visión global del sistema, presentada en la Fig. 5, refleja el flujo que sigue la aplicación, desde que es generada por los puntos de acceso, hasta que es consumida por el usuario final del interfaz gráfico.

### B. Pasarela de datos

El correcto funcionamiento de toda la aplicación desarrollada depende de la gestión de los datos compartidos entre las capas, para lo que se define la llamada pasarela de datos, diseñada a partir de un servicio que implementa FloodLight y que es la base de datos intermedia que alberga la información necesaria para la aplicación web.

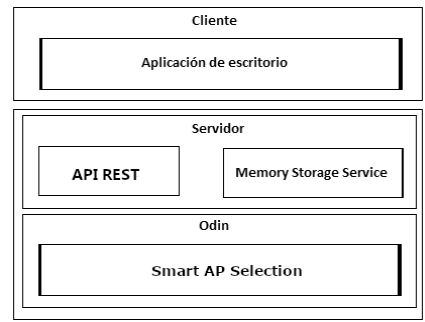

Fig. 4. Arquitectura del sistema desarrollado.

La pasarela incluye una función que inicia las tablas NoSql al inicio del proceso de la aplicación, antes de iniciar los escaneos periódicos. Posteriormente, las diferentes funciones atacarán estas tablas usando operaciones CRUD *(Create, Read, Update and Delete)*. El servicio puede almacenar objetos Java en las tablas.

La información que genera cada uno de los procesos de *Smart AP Selection*, así como la información obtenida directamente de los puntos de acceso, se vuelca sobre la pasarela, donde los datos actualizados están disponibles para los clientes que la requieran.

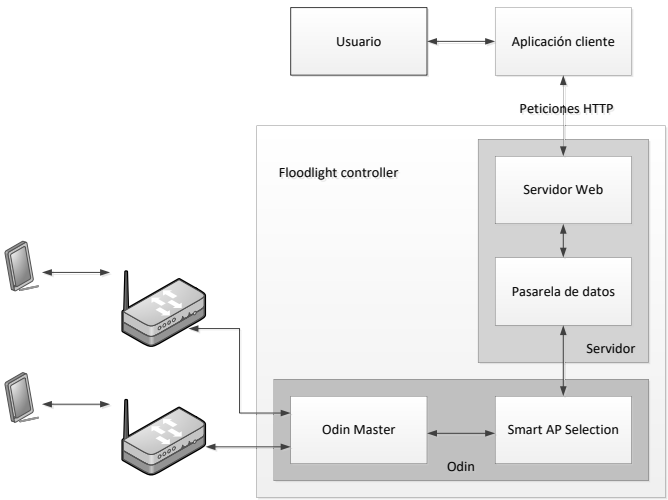

Fig. 5. Visión global de los elementos del sistema.

Para el traspaso seguro de información entre la aplicación web y *Smart AP Selection* se cuenta con el servicio de memoria presente en el sistema Floodlight. Este servicio actúa como una base de datos NoSql, dando persistencia a los datos mientras el sistema esté en ejecución. Su estilo NoSql requiere de la creación de entidades donde se guarde la información que interesa al usuario. Dichas entidades, al generarse en memoria cacheada, deben ser creadas en cada ejecución del programa. Las entidades creadas se almacenarán en forma de tablas NoSql de objetos Java.

Se busca una correcta interacción de la aplicación *Smart AP Selection* con el servicio que actúa de pasarela. Por tanto, durante el periodo entre la finalización del proceso y el reinicio del escaneo periódico, se comprueba si el cliente ha realizado alguna petición para modificar los parámetros de la aplicación y, si es así, se realiza la modificación antes de que el bucle principal se reinicie (Fig. 6).

Los pies de las figuras y de las tablas deben seguir el formato mostrado bajo la Fig. 1 y sobre la tabla I. Si es posible, utilice un formato vectorial (como EPS o PDF) para representar diagramas. Los formatos de tipo *raster* (como PNG o JPG) suelen generar ficheros muy grandes y pueden perder calidad al ampliarlos.

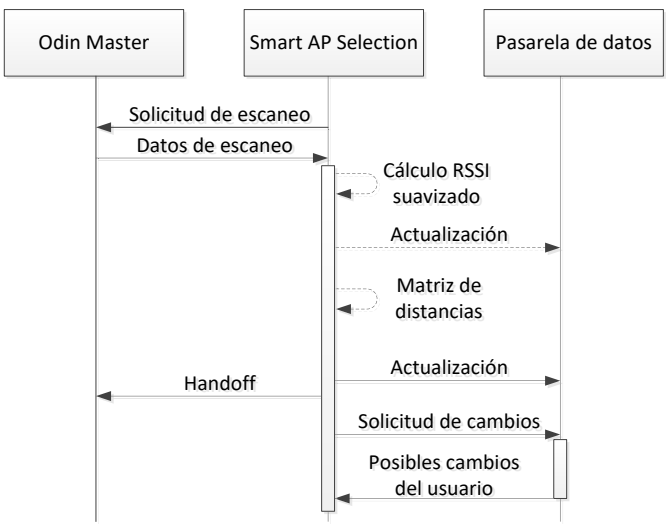

Fig. 6. Diagrama de secuencia de la información.

## C. Servicio web

El servicio web, juntamente con la pasarela, actúa como intermediario entre el cliente y la aplicación *Smart AP Selection* independizando un servicio del otro. El servicio web dispone de diferentes *endpoints*:

- Obtención de todos los clientes que están actualmente conectados o lo hayan estado en algún momento.
- Estaciones que están actualmente conectadas.
- Agentes que actualmente están conectados.
- Se indica al sistema que un AP debe cambiar el canal donde está emitiendo. Se debe indicar la dirección IPv4 del AP y el canal al cual se desea cambiar.
- Solicitud del *handoff* de una estación de manera manual. El *handoff* requerirá de dos parámetros: la dirección IP del AP al que se desea mover la estación, y la dirección física MAC que identifica al cliente.
- Petición que permita que el usuario pueda solicitar un escaneo de un canal por parte del AP. Los AP pueden escanear en cualquier momento, si no se da el caso que se encuentren ya escaneando. En ese caso, este *endpoint* rescata los datos del último escaneo e informa de ellos al usuario.
- Obtención de los parámetros de la aplicación.
- Solicitud de modificación de algún parámetro.
- Envío de la matriz de potencias, que se mostrará al usuario.

Todos los *endpoint* propuestos responden mediante objetos JSON. De esta manera se estandariza la comunicación entre los servicios.

## D. Cliente

El cliente se ha desarrollado como una aplicación de escritorio que mediante peticiones HTTP que obtiene la información que requiere de la aplicación *Smart AP Selection*. Con el fin de realizar una capa mantenible y robusta, se optó por un modelo modelo-vista-controlador [8] (MVC) (Fig. 7). Dicho modelo subdivide una aplicación en tres partes principales, con el objetivo de separar la lógica de negocio y sus datos, de la capa donde interactúa el usuario y todos los controladores encargados de responder a los eventos generados. El acceso a los datos se realiza mediante un conjunto de peticiones al servicio HTTP dispuesto por el controlador.

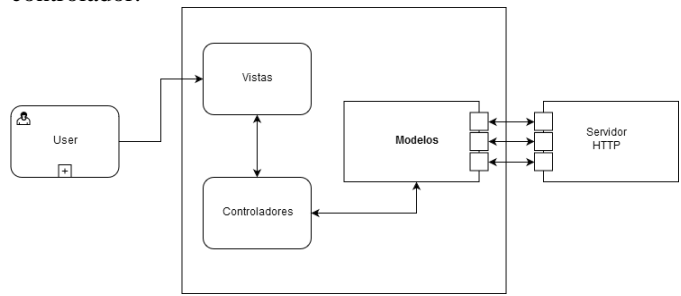

Fig. 7. Modelo MVC de la aplicación de escritorio.

Se han diseñado las diferentes vistas de la aplicación, presentando los datos de la red de una manera amigable y en tiempo real. Se diseñan tres vistas diferentes acordes a los diferentes casos de uso planteados: *network*, *clientes* y *estadísticas*. Cada una tiene el objetivo de mostrar de una manera agradable y útil la compleja información que produce el sistema.

Para la pantalla *network* (Fig. 8) se propone un gráfico piramidal, con el controlador arriba, los diferentes puntos de acceso un nivel por debajo, las estaciones (STA) otro nivel por debajo, representando las conexiones mediante líneas continuas entre los nodos. El gráfico indicará el nivel de señal (RSSI) de cada estación. De la misma manera, se utilizarán diferentes colores en la línea (variando de verde a rojo), según sea la calidad de la señal.

Tras un estudio de los requerimientos y de las posibles tecnologías compatibles con Java, se optó por la librería GraphStream (http://graphstream-project.org/), que permite mostrar grandes cantidades de datos mediante grafos en tiempo real.

En la vista *network*, el usuario puede ver el estado de la red en tiempo real, incluyendo los puntos de acceso, estaciones conectadas y su estado de conexión, así como una matriz de distancias (o atenuación) entre las estaciones conectadas y los puntos de acceso. Además, necesita de datos actualizados de manera constante, por tanto, se desarrolló una tarea que, de manera periódica, solicita los datos del último escaneo, obteniendo los últimos cambios que se han producido en la red. Esta tarea se realiza con *Timeline*, componente de la librería de JavaFX, y se ejecuta periódicamente, según se indique en el parámetro *intervalo de escaneo*. Además, desde esta vista el usuario podrá modificar algunos de los parámetros de la aplicación *Smart AP Selection*.

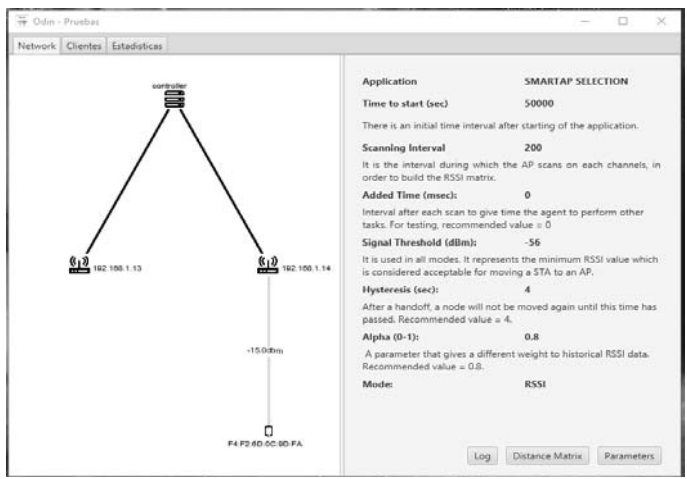

Fig. 8. Captura de pantalla de la vista general.

La matriz de distancias entre puntos de acceso y clientes (Fig. 9) se muestra con los mismos colores que se muestran en el gráfico inicial, indicando de esta manera al usuario la calidad de las diferentes conexiones entre puntos de acceso y estaciones.

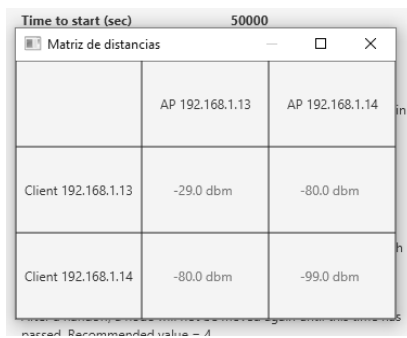

Fig. 9. Captura de pantalla del pop up con la matriz de atenuación

La vista de *estadísticas del sistema* (Fig. 10) presenta de manera gráfica los datos del último escaneo. Estas son las gráficas que se han considerado útiles:

- El tiempo en el aire de todas las estaciones conectadas a un punto de acceso escogido.
- El RSSI suavizado de cada estación para un punto de acceso escogido.
- El número de paquetes de cada estación para un punto de acceso escogido.

Estos tres parámetros permiten al administrador de la red comprobar que el algoritmo de balanceo está funcionando correctamente. Si una estación ocupa mucho tiempo en el aire, enviando pocos paquetes y con un RSSI bajo, estará probablemente posicionada en un punto de acceso no óptimo, recibiendo peor señal de la que podría tener en otro punto de acceso. En el caso opuesto, una estación con buena señal debería ser capaz de enviar grandes cantidades de paquetes ocupando un menor tiempo de aire.

En la vista se muestran los datos del último escaneo realizado por los puntos de acceso. Durante un periodo de tiempo los agentes miden el número de paquetes que envía una estación, el tiempo que ha tardado en enviar ese número de paquetes, y el nivel de señal con la cual se envían dichos paquetes.

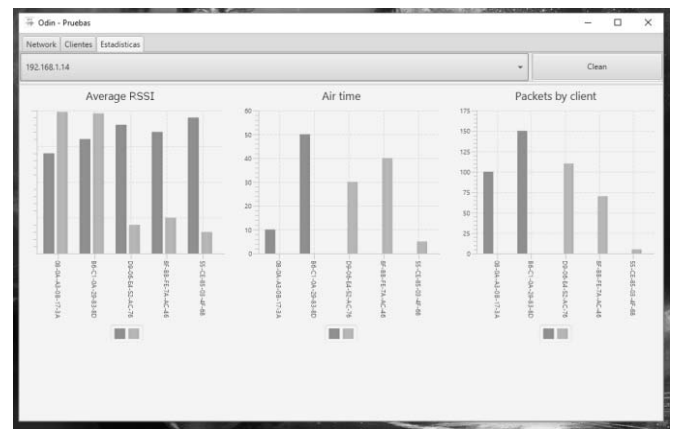

Fig. 10. Gráficas de estadísticas de la red Wi-Fi.

Finalmente, la Fig. 11 muestra la pantalla *clientes* y en particular una captura de pantalla que corresponde a la "solicitud de handoff" que el gestor del sistema puede realizar, para solicitar que una estación cambie a otro AP.

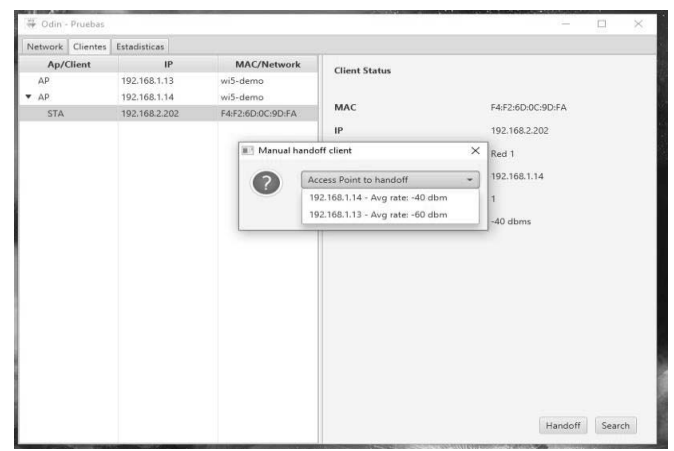

Fig. 11. Ventana emergente de solicitud de handoff.

## IV. Conclusiones

En este trabajo se ha presentado el desarrollo de una interfaz gráfica para la gestión de una WLAN que integra un conjunto de AP Wi-Fi coordinados. Se ha presentado el sistema, así como la aplicación Smart AP Selection, que se encarga de realizar un balanceo de carga y una gestión de la movilidad. La aplicación es capaz de obtener y mostrar la información almacenada en un sistema complejo como es Wi-5, y se muestra ahora de manera amigable para el gestor de la red. Además, la aplicación permite la gestión remota del sistema, ajustando sus parámetros o realizando traspasos entre puntos de acceso.

El trabajo realizado con las diferentes tecnologías usadas (*Java*, *JavaFx*, *FloodLight*, entre otras) y la interrelación entre ellas ha concluido en un sistema robusto, capaz de cumplir con los objetivos y requisitos requeridos, obteniendo una solución usable, atractiva, y mantenible, y que puede ser usado como base para siguientes trabajos, ya que ha sido diseñado con este objetivo.

## Agradecimientos